\documentclass[onecolumn]{emulateapj}
\usepackage{amsmath}
\shortauthors{Allred et al.}
\shorttitle{RHD Models of Emission from Solar Flares}
\begin{document}

\title{Radiative Hydrodynamic Models of the Optical and
Ultraviolet Emission from Solar Flares}
\author{Joel C. Allred\altaffilmark{1}, Suzanne L. Hawley\altaffilmark{2},
William P. Abbett\altaffilmark{3}, and Mats Carlsson\altaffilmark{4}}
\altaffiltext{1}{University of Washington, Department of
   Physics, Box 351560, Seattle, WA 98195.}
\altaffiltext{2}{University of Washington, Department of
   Astronomy, Box 351580, Seattle, WA 98195.}
\altaffiltext{3}{Space Sciences Laboratory, University of California,
   Berkeley, California, 94720}
\altaffiltext{4}{Institute of Theoretical Astrophysics, University
   of Oslo, P.O. Box 1029, Blindern, N-0315 Oslo, Norway}

% The abstract environment prints out
% horizontal rules for the journal's editorial staff to type the
% received and accepted dates

\begin{abstract}
We report on radiative hydrodynamic simulations of moderate and
strong solar flares. The flares were simulated by calculating the
atmospheric response to a beam of non-thermal electrons injected
at the apex of a one-dimensional closed coronal loop, and include
heating from thermal soft X-ray, extreme ultraviolet and
ultraviolet (XEUV) emission. The equations of radiative transfer
and statistical equilibrium were treated in non-LTE and solved
for numerous transitions of hydrogen, helium, and Ca {\sc ii}
allowing the calculation of detailed line profiles and continuum
emission. This work improves upon previous simulations by
incorporating more realistic non-thermal electron beam models and
includes a more rigorous model of thermal XEUV heating.
We find XEUV backwarming contributes less than 10\% of
the heating, even in strong flares. The simulations show elevated
coronal and transition region densities resulting in dramatic
increases in line and continuum emission in both the UV and
optical regions. The optical continuum reaches a peak increase of
several percent which is consistent with enhancements observed in
solar white light flares. For a moderate flare ($\sim$M-class),
the dynamics are characterized by a long gentle phase of near
balance between flare heating and radiative cooling, followed by
an explosive phase with beam heating dominating over cooling and
characterized by strong hydrodynamic waves. For a strong flare
($\sim$X-class), the gentle phase is much shorter, and we
speculate that for even stronger flares the gentle phase may be
essentially non-existent. During the explosive phase, synthetic
profiles for lines formed in the upper chromosphere and
transition region show blue shifts corresponding to a plasma
velocity of $\sim$120 km s$^{-1}$, and lines formed in the lower
chromosphere show red shifts of $\sim$40 km s$^{-1}$.
\end{abstract}

\keywords{methods: numerical --- radiative transfer --- Sun: atmosphere --- Sun: flares}

% Next comes the main body of the paper.
% In the first two sections, you should notice the use of the LaTeX
% \cite command to identify citations.  The tags are tied to the
% citations in the references section.  Please go to the LaTeX
% manual for a complete description of the \cite-\bibitem mechanism.

\section{Introduction}
During a solar flare, large quantities of energy are transferred
between the corona and chromosphere through thermal conduction,
non-thermal particle beams, radiation transport and mass motions.
Any attempt to accurately model the optical and UV radiation
produced during a flare must incorporate these methods of energy
transfer. There has been considerable effort put into modeling the
atmospheric response to heating from non-thermal particle beams.
\citet{b73} and \citet{e78} derived an analytic expression
describing electron energy deposition through Coulomb collisions.
This expression has been employed in a number of flare simulations
\citep{rc83,m84, ne84, fcm85, p87, mel89, hf94, ah99}.
Understanding how energy is transferred by mass motions and
radiative transfer requires solving the equations of radiation
hydrodynamics. Many hydrodynamical simulations have focused on the
upper atmosphere where emission is optically thin, and radiative
transfer is relatively straightforward. Notable investigations of
this type are \citet{c83, c84, mel89, pr93, rp95}. However, we
seek to model optical emission which originates in the
chromosphere where radiation is neither optically thin nor in LTE
with the surrounding plasma. Therefore, we must employ the full
non-LTE formulation of radiative transfer.

The simulations of \citet{fcm85} were among the first to model the
radiative hydrodynamic response in a coronal loop to a beam of
non-thermal electrons.  However, their models were limited in that
radiative transfer was solved using an escape probability method.
The models of \citet{hf94} incorporated non-thermal beam heating
into the radiative transfer code, MULTI \citep{c86}. This had the
advantage of efficiently solving the equation of radiative
transfer, but did not include detailed hydrodynamics, and
therefore could not simulate the effects of mass motions on the
energy balance or on the line profiles.  The models of
\citet{ah99} (hereafter, AH99) incorporated both hydrodynamics and
radiative transfer, and are therefore well suited for
understanding the optical emission produced during flares.  In
this paper we extend the models of AH99 by more accurately
modeling the XEUV radiation and electron beam heating.

In section 2 we describe the method by which the radiative
hydrodynamic equations are solved. We also describe the procedure
by which we generate the initial atmosphere and our new methods
for incorporating soft X-ray and electron beam heating into the
simulations.  In section 3 we discuss the dynamics of our
simulations and present detailed line and continuum profiles. The
limitations of these models and future improvements are described
in section 4, and in section 5 we present our conclusions.

\section{Method of Solution}
The basic method of solution is described in detail in AH99 and
\citet{a98}. Here we briefly outline the procedure.  The equations
of hydrodynamics, population conservation and radiative transfer
are solved simultaneously on a one-dimensional adaptive grid
\citep{dd87} using the RADYN code of \citet{cs94,cs95,cs97}. The
code has been modified to include the effects of flare heating
from a non-thermal electron beam, heating from high temperature
soft X-ray, extreme ultraviolet and ultraviolet emission
(hereafter, we refer to these collectively as XEUV emission) and
optically thin cooling resulting from bremsstrahlung and
collisionally excited metal transitions. Hydrodynamic effects due
to gravity, thermal conduction and compressional viscosity are
included as described in AH99.  We have additionally modified the
code to include the double power-law electron beam energy
distributions recently observed in solar flares with the RHESSI
satellite \citep{h03}. We incorporate the effects of XEUV heating
from a large number of high temperature lines using results from
the CHIANTI \citep{d97,y03} and ATOMDB databases \citep{sblr01}.

We obtain the non-LTE solution to the population equations for a
six level with continuum hydrogen atom, nine level with continuum
helium atom and six level with continuum Ca {\sc ii} atom using
the technique of \citet{sc85}. The bound-bound and bound-free
transitions which are calculated in detail are listed in
Tables~\ref{table:bb}~\&~\ref{table:bf}.  All lines are calculated
assuming complete redistribution, but for the Lyman transitions
the effects of partial redistribution are approximated by
truncating the line profiles at 10 Doppler widths.  Other
transitions are included as background opacity in LTE.  The
opacities are computed from the Uppsala package of \citet{g73}. We
include optically thin radiative cooling from bremsstrahlung and
coronal metal transitions using emissivities from the ATOMDB
database. The equations are solved on an adaptive grid using 191
grid points in depth, 5 in angle and up to 100 in frequency for
each transition. We have found it necessary to employ an adaptive
grid in order to resolve large shocks which propagate through the
atmosphere.

\subsection{The Preflare Atmosphere}
A magnetic flux tube is typically modeled as a semi-circular loop
with footpoints embedded in the photosphere and apex in the
corona.  By symmetry we need only consider one leg of the loop
which we take to be a cylinder of constant cross section. In this
approximation the atmospheric height, $z$, is the only spatial
degree of freedom.  We use the PF2 model of AH99 as the preflare
atmosphere. The PF2 atmosphere was generated by adding a
transition region and corona to the model atmosphere of
\citet{cs97}. To resolve the steep gradients present in the
transition region PF2 required 191 grid points.  Constant
non-radiative quiescent heating was applied to grid zones with
photospheric column mass (i.e. column mass greater than 7.6 g
cm$^{-2}$) to balance the energy losses in the photosphere. The
upper boundary was set at $10^4$ km and held at $10^6$ K. With
these boundary conditions and no external sources of heating, the
atmosphere was allowed to relax to a state of hydrodynamic
equilibrium.

\subsection{Electron Beam Heating \label{sec:ebeam}}
In the standard flare reconnection model, electrons located near
the loop apex are accelerated to high energies and travel downward
through the atmosphere depositing energy and producing hard X-ray
bremsstrahlung radiation. For the simulations presented here, we
assume that the electron beam is the primary source of flare
heating in the lower solar atmosphere.  Recently, \citet{h03}
analyzed hard X-ray spectra observed with RHESSI to obtain the
injected electron energy spectrum at 20 s intervals throughout the
23 July 2002 X-class solar flare. The injected electrons are found
to have a double power law energy distribution of the form
\begin{equation}
\label{eqn:dpl}
 F_0(E_0) = \frac{ \mathcal{F} (\delta_u-2) (\delta_l-2)}
{E_c^2 \left( (\delta_u-2) - \left( \frac{E_B}{E_c}
\right)^{2-\delta_l} (\delta_u-\delta_l) \right) } \left\{
\begin{array}{ll}
\left( \frac{E_0}{E_c} \right) ^{-\delta_l} & \textrm{for
$E_0<E_B$} \\\left(\frac{E_B}{E_c} \right )^{\delta_u-\delta_l}
\left(\frac{E_0}{E_c} \right) ^{-\delta_u} & \textrm{for
$E_0>E_B$}\end{array} \right.
\end{equation}
where $\mathcal{F}$ is the electron energy flux that
enters the magnetic loop; $E_c$ is the cutoff energy below which
the X-ray emission is assumed to be thermal; and $E_B$ is the
break
energy where the distribution shifts from spectral index
$\delta_l$ to $\delta_u$.  Using Equation~\ref{eqn:dpl} we
calculated
the energy deposition rate as a function of column depth using the
method of \citet{e78,e81} which assumes that the energetic
particles are slowed and deposit their heat through Coulomb
interactions.  To account for the changing partial ionization of
hydrogen we have adopted the treatment of \citet{hf94}.  The
resulting heating rate as a function of column depth is,
\begin{align} \label{eqn:qe}
Q_e(N) = \frac{K \mathcal{F}}{2 \mu_0 E_c^2}\gamma(N)
\frac{(\delta_u-2) (\delta_l-2)}{\left( \frac{E_B}{E_c}
\right)^{2-\delta_l}
(\delta_l-\delta_u)+(\delta_u-2)} & \nonumber \\
 \left\{ \left( \frac{N^*(N)}{N^*_c}
\right)^{-\frac{\delta_l}{2}} \left[ B_{x_c}
 \left( \frac{\delta_l}{2},\frac{1}{3} \right)
-B_{x_B} \left( \frac{\delta_l}{2},\frac{1}{3} \right) \right]
+ \right. \nonumber \\
\left.
\left( \frac{N^*(N)}{N^*_c} \right)^{-\frac{\delta_u}{2}} \left(
\frac{E_B}{E_c} \right)^{\delta_u-\delta_l} B_{x_B} \left(
\frac{\delta_u}{2},\frac{1}{3} \right) \right\}
\end{align}
where $K=2 \pi e^4$, $\gamma(N)=x \Lambda + (1-x) \Lambda'$, $x$
is the hydrogen ionization fraction, $\Lambda$ and $\Lambda'$ are
the Coulomb logarithms defined in \citet{e78}, and $\mu_0$ is the
cosine of the pitch angle of the beam; $\mu_0 =1 $ in our
simulations.  The quantity $B_{x_c}(\delta/2, 1/3)$ is the
incomplete beta function for $x_c= N/N_c$ where $N_c = \mu_0
E_c^2/3 K \gamma$ is the maximum column depth penetration of a
beam with cutoff energy $E_c$.  $x_B$ is defined similarly for
$E_B$. $N^*(N)$ is the equivalent column depth in a fully ionized
plasma defined by $N^*(N)=\int_{0}^{N} \gamma(N')/\Lambda dN'$
and $N^*_c$ is the cutoff depth in a fully ionized plasma.

The parameters $\delta_l$, $\delta_u$, $\mathcal{F}$, $E_c$ and
$E_B$ are obtained by fitting the hard X-ray spectra observed
during a flare to models for thick target bremsstrahlung as
described in \citet{h03}. In a future paper, we will use the
time-dependent beam parameters they have obtained to simulate the
23 July 2002 solar flare.  In this paper, we take $\delta_l$,
$\delta_u$, $E_c$, and $E_B$ to be constant with values of 3.0,
4.0, 37 keV and 105 keV, respectively, corresponding to the peak
of the 23 July 2002 flare.

\begin{figure}
\epsscale{.6}
    \plotone{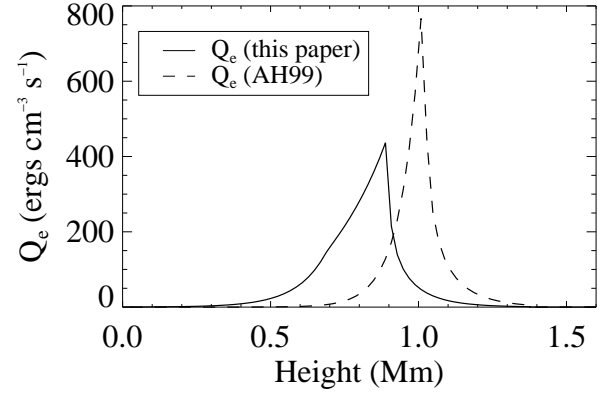}
    \caption{Electron beam heating rates in the preflare
atmosphere for the F10 flare.  The solid line shows the heating
rate used in this paper (defined in Eq.~\ref{eqn:qe}) and the
dashed line shows the initial beam heating rate of AH99.  Our use
of recently measured spectral indices and cutoff energy for the
beam as described in \S~\ref{sec:ebeam}, results in extended beam
heating that penetrates deeper in the atmosphere than the models
described in AH99.}
    \label{fig:qe}
\end{figure}

Figure~\ref{fig:qe} shows the initial energy deposition rate of
the electron beam and compares it with the initial deposition
rate from AH99.  AH99 used a single power law electron energy
spectrum with a spectral index, $\delta= 5$, and a cutoff energy,
$E_c=20$ keV.  The lower spectral indices and the higher energy
cutoff we employ result in energy deposition that extends
deeper in the atmosphere and has a broader spatial extent.

Non-thermal electrons affect the atmosphere not only by energy
deposition but also through direct collisional ionization. We
estimate collisional ionization due to non-thermal electrons
using the technique of \citet{rc83}.  They find the non-thermal
collisional ionization rate from the ground state to be
\begin{equation}\label{eqn:ntci}
    C_{nt} = 3.78 \times 10^9 \frac{Q_e^{(n)}}{(1-x) n_H}
\end{equation}
where $n_H$ is the hydrogen number density and $Q_e^{(n)}$ is the
beam energy deposition rate due to collisions with neutral
hydrogen and is related to the total energy deposition rate
defined in Equation~\ref{eqn:qe} by $Q_e^{(n)} = Q_e (1-x)
\Lambda'/\gamma$.  They also find that ionizations from the
ground state dominate, so we neglect non-thermal collisional
ionizations from excited states.

\subsection{Soft X-Ray, EUV and UV (XEUV) Heating
\label{sec:xray}}
Mass motions during flares raise the density in the transition
region and corona which can greatly increase the emissivity of
XEUV photons from these regions. The outward directed photons we
detect as elevated emission, while the downward directed photons
cause ``backwarming'': heating of the lower atmosphere due to
increased photoionizations.  We have used the ATOMDB database to
determine the thermal volume monochromatic emissivity as a
function of wavelength and temperature. We include emissivities
calculated for approximately 34,000 transitions at 37 temperature
points ranging from $10^4$ K to $10^7$ K.  The transitions have
been gathered into 14 wavelength bins ranging from 1 \AA$\,$ to
2500 \AA$\,$. Table~\ref{table:xbins} lists the range and
emissivity-weighted central wavelengths of each bin. When
calculating these emissivities, we have taken care not to include
transitions that are already solved in detail in RADYN (cf.
Table~\ref{table:bb}).

Since the XEUV emission produced in low-temperature,
optically-thick layers is negligible, we assume that all such
emission originates from the higher temperature, optically-thin
layers.  The XEUV flux incident on a layer with optical depth
$\tau_\nu$ is given by,
\begin{equation}\label{eq:fxpp}
   F_\nu(\tau_\nu)=2\pi\int_{\tau_\nu}^\infty S_{\nu}
   (t_\nu) E(t_\nu-\tau_\nu)\,dt_\nu - 2\pi\int_0^{\tau_\nu}
   S_{\nu}(t_\nu) E(\tau_\nu-t_\nu)\,dt_\nu \;,
\end{equation}
where $S_\nu$ is the source function, given by the ratio of the
emissivity to the linear extinction coefficient.  For layers
close to the layer under consideration ($r < d/\sqrt{ 8}$, where
$r$ is the separation between layers and $d$ is the loop
cross-sectional diameter), $E$ in the above integral is $E_2$,
the second exponential integral. However, in order to avoid
overestimating the emission from distant layers ($r > d/\sqrt{
8}$) we follow \citet{gan90,hf94}; and AH99 in treating the
distant layers as point sources.  In that case $E$ is given by,
\begin{equation}
E(\tau)=\frac{\mu' d^2}{32\pi r^2} e^{-\tau/\mu'}
\end{equation}
where $\mu'$ is defined in Equation 10 of AH99.

The XEUV photons heat the lower atmosphere through
photoionization. The average scale height of our model atmosphere
is greater than the thermalization depth scale of \citet{hen77},
so we assume that the energy of the photoionized electrons is
transformed entirely to heat in each atmospheric layer.  The
volumetric heating rate is given by,
\begin{equation}\label{eq:xheat}
   Q_{XEUV}=\int_{\nu} \frac{F_\nu}{h\nu} \left[ \sum_j
n_j(h\nu-X_j)\sigma_{j\nu} \right] d\nu\;,
\end{equation}
where $n_j$ is the number density for ion $j$, $X_j$ is the
ionization potential, $\sigma_{j\nu}$ is the photoionization cross
section, and the sum is over approximately 150 ions. The cross
sections and number densities were obtained from the CHIANTI
database.

\begin{figure}
\epsscale{.6}
    \plotone{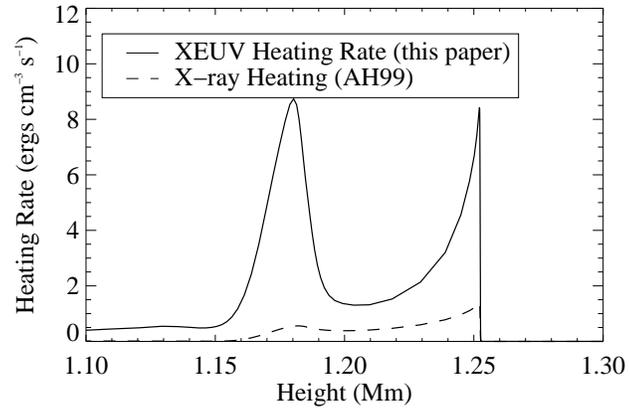}
    \caption{The XEUV heating rate as a function of atmospheric
    height compared with the soft X-ray heating rate of AH99 at
a comparable time in the late stages of a moderate flare.}
    \label{fig:compxray}
\end{figure}

Our method for calculating the XEUV heating is similar to the
soft X-ray heating of AH99, but differs in the extent of the
atomic transitions included. AH99 considered emission only
between 1 \AA{} and 250 \AA{} divided into seven bins, while we
include 14 bins in the range 1 \AA{} to 2500 \AA{}, i.e.
including substantial EUV and UV heating.  For example,
C~\textsc{iii} $\lambda$977 \AA{}, which was not included in
AH99, is a significant contributor in our backwarming
calculations. Figure~\ref{fig:compxray} compares our XEUV heating
rate with that of AH99 at a comparable time late in a moderate
flare. Our XEUV heating rate is approximately 10 times larger
than the rate in AH99.

\section{Flare Simulations}
We carried out two flare simulations, corresponding to moderate
beam heating with an electron flux, $\mathcal{F}=10^{10}$ ergs
cm$^{-2}$ s$^{-1}$ (the F10 flare) and strong beam heating, with
$\mathcal{F}= 10^{11}$ ergs cm$^{-2}$ s$^{-1}$ (the F11 flare).
The F10 flare was evolved for 226 s and the F11 flare for 15 s.
Note that these simulations are comparable to the F10 and F11
flares in AH99 which evolved for 70 s and 4 s, respectively.  In
the following sections we discuss the dynamics and emission
resulting from these two flare simulations.

\subsection{F10 Flare Dynamics}
\begin{figure}
\epsscale{1}
    \plotone{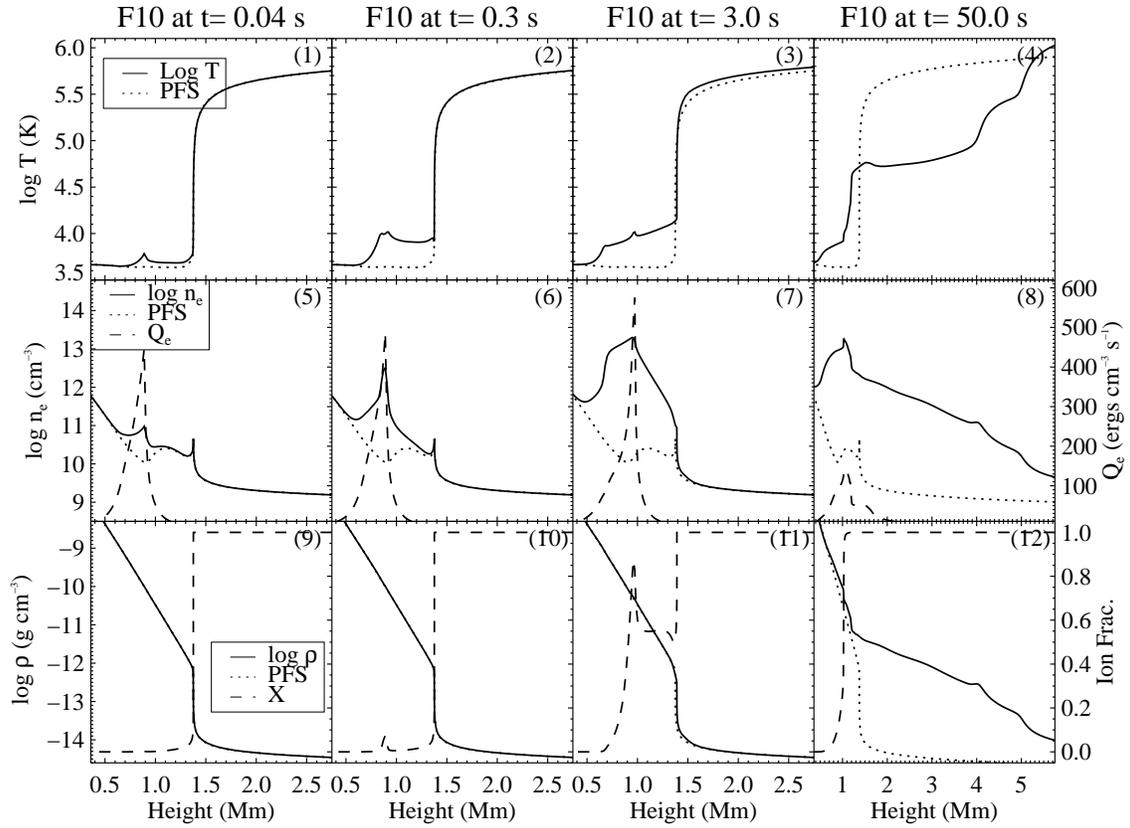}
    \caption{The solar atmosphere at four times during the
gentle phase of the F10 flare.  The top row shows the
log of the temperature, $T$ as a function of height compared with
the preflare state (PFS).  In the middle row the electron
density, $n_e$ (left axis) and beam heating rate, $Q_e$ (right
axis) are plotted.  The bottom row shows the mass density, $\rho$
(left axis) and hydrogen ionization fraction, $X$, (right axis).
Note the change in scale of the horizontal axis in the last
column.}
\label{fig:f10gen}
\end{figure}

Figure~\ref{fig:f10gen} shows the evolution of the temperature
and density stratification along with the electron beam energy
deposition rate, the ionization fraction and the electron density
during the F10 flare. Initially the beam penetrates into the
chromosphere to a depth of 0.89 Mm above the solar surface
(defined as $z=0$ at $\tau_{5000}  = 1$).  The upper chromosphere
rapidly heats but plateaus at a temperature of about $10^4$ K
after 0.3 s as a result of hydrogen radiative cooling (see panel
2 of Fig.~\ref{fig:f10gen}). The initial beam impact causes the
pressure to increase and a strong wave begins to propagate
upward. The dynamics proceed slowly for the next three seconds,
as much of the incident beam energy is radiated away.  By 3.0 s,
a significant fraction of the hydrogen in the region of beam
energy deposition has been ionized and is no longer capable of
radiating the beam energy, and so the temperature again quickly
rises (see the hydrogen ionization fraction in panel 11 of
Fig.~\ref{fig:f10gen}). Over the next several seconds, the
increasing temperature causes He~\textsc{i} in the atmospheric
range of 1.0 to 1.4 Mm to become ionized to He~\textsc{ii}. The
cooling increases through He recombination and the
He~\textsc{ii}~$\lambda$304 line, and the radiative cooling again
balances the beam and XEUV heating. When this balance is reached,
a relatively gentle phase of the evolution begins. Meanwhile, the
wave which resulted from the initial impact of the beam has been
traveling through the atmosphere, bringing material from the
chromosphere into the transition region and corona.  By 50 s the
initial wave has reached the top of the flux tube.  Material
pushed by the wave has increased the apex density by a factor of
three and raised the temperature to about 2 MK. The increased
coronal density causes the transition region to become
much more extended in height than in the preflare atmosphere
(panel 4 of Figure~\ref{fig:f10gen}). In addition, with
elevated coronal density the atmosphere becomes more effective at
stopping the beam, which therefore deposits its energy
progressively higher in the atmosphere. Column 4 of
Figure~\ref{fig:f10gen} (panels 4, 8, 12) illustrates these
effects at a representative time during the gentle phase.

After F10 has evolved for about 73 s, most of the He~\textsc{ii}
in the range of 2.0 to 2.3 Mm has been ionized, and the radiative
cooling again decreases below the flare heating.  This is
illustrated in the first column of Figure~\ref{fig:f10exp}. An
explosive increase in temperature results and creates a
supersonic wave which pushes material away from its front and
forms a lower density (though still much higher than the preflare
density) high temperature bubble. The second column of
Figure~\ref{fig:f10exp} (panels 2, 6, 10) shows these effects at
85 s. Meanwhile, more material has been evaporated into the
corona and the loop apex density has increased approximately an
order of magnitude over the preflare density (panel 10).  For the
next 150 s the bubble expands as the explosive wave propagates
through the atmosphere, until the wave reaches the boundary of
the flux tube. The atmosphere attains a steady state, with a
hotter, denser corona, and a transition region that is closer in
height to the photosphere than in the preflare atmosphere (panels
4, 8, 12). The final apex temperature is 5.2 MK, the electron
density is $5.9 \times 10^{10}$ cm$^{-3}$, and the mass density
has increased by a factor of 70 reaching $1.2 \times 10^{-13}$ g
cm$^{-3}$. This temperature and density are somewhat lower than
coronal values often observed for M-class flares. Although the
XEUV heating rate would be larger from an even hotter, denser
corona, it would still be much smaller than the beam
heating (see below), and therefore our predictions of the optical
and UV emission from the lower atmosphere are not affected.  For
example, an atmosphere with an apex temperature of 20 MK and
electron density of $5 \times 10^{11}$ cm$^{-3}$ gives an XEUV
heating rate which is five times larger than ours, but still 50
times smaller than the beam heating rate.

A major difference between the models presented here and those of
AH99 is the time at which the explosive phase begins.  The
explosive phase of the F10 model of AH99 begins after about 27 s
of impulsive heating compared to 73 s for our F10 model.  The
reason is that the lower spectral indices we employ
result in broader electron beam energy deposition.  With less
concentrated heating, the atmosphere takes longer to ionize
resulting in a longer delay before the onset of the explosive
phase.

\subsection{F11 Flare Dynamics}
The initial evolution of the F11 flare is similar to F10 but
proceeds much faster.  Figure~\ref{fig:f11atmos} shows the
atmosphere at several times during this flare.  The beam rapidly
heats the atmosphere in the region of impact to about 13,000 K,
but plateaus after 0.04 s because hydrogen radiative cooling
balances the beam heating. As in the F10, case a high velocity
wave begins to propagate upward from the initial beam impact. By
0.18 s, most of the hydrogen is ionized and the atmosphere again
rapidly heats, but plateaus at just over 50,000 K as a result of
He~\textsc{ii} radiative cooling (first column of
Fig.~\ref{fig:f11atmos}). By 1.0 s, helium is completely ionized
and the radiative cooling is insufficient to balance the beam
heating, causing the atmosphere to explosively heat (second
column of Fig.~\ref{fig:f11atmos}). A high-temperature,
low-density bubble forms in a fashion similar to the F10 flare,
but at a location much lower in the atmosphere than in the F10
case. Since the atmosphere heats so much more quickly in the F11
flare, the initial wave resulting from beam impact has not had
time to evaporate much material into the corona. Therefore, the
beam is still depositing most of its energy in the lower
chromosphere when the explosive phase starts. The explosive
bubble forms at about 1 Mm (panel 2 of Fig.~\ref{fig:f11atmos}).
A supersonic wave forms as a result of the explosion, and it is
this wave which is primarily responsible for depositing
chromospheric material into the corona, in contrast to the F10
case where the material was carried primarily by the initial
wave.  It is likely that for flares larger than F11, the onset of
the explosive phase will be extremely fast (i.e. hydrogen and
helium will be ionized almost immediately). For very large
flares, the explosive phase will essentially begin simultaneously
with the beam heating, with no observable lag.

\citet{hf94} speculated that XEUV emission could provide
significant heating below the temperature minimum region, which
might explain the increased continuum emission produced in white
light flares. However, in both F10 and F11, we find that the XEUV
heating does not penetrate more deeply than the electron beam,
nor does it contribute significantly to the energy deposition
rate. Despite an order of magnitude more heating than in AH99,
XEUV heating accounts for less than 1\% of the total energy
deposited in the F10 flare and about 7\% in the F11 flare.

\subsection{Line Emission}
\begin{figure}
    \plotone{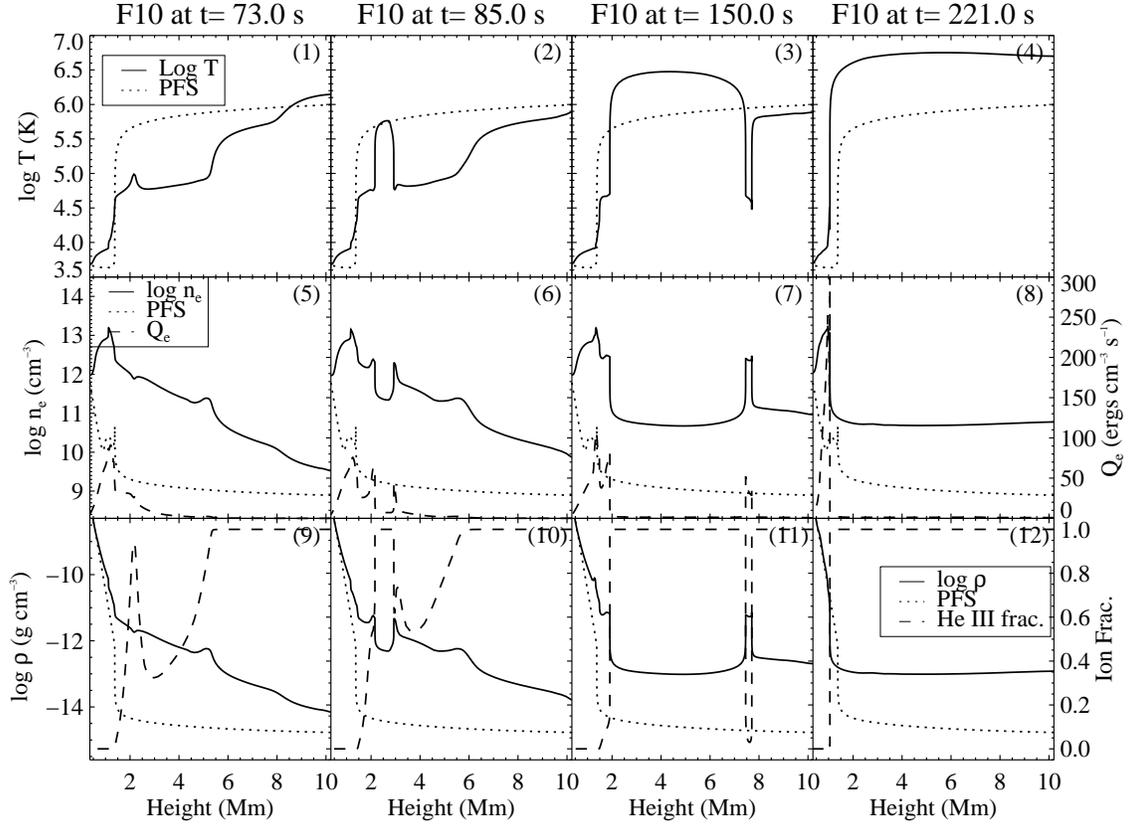}
    \caption{The F10 solar atmosphere at four times during the
    explosive phase.  The quantities plotted are identical to Fig.~\ref{fig:f10gen}
    except that the He \textsc{iii} fraction is plotted in the bottom row
    rather the hydrogen ionization fraction.}
    \label{fig:f10exp}
\end{figure}

\begin{figure}
    \plotone{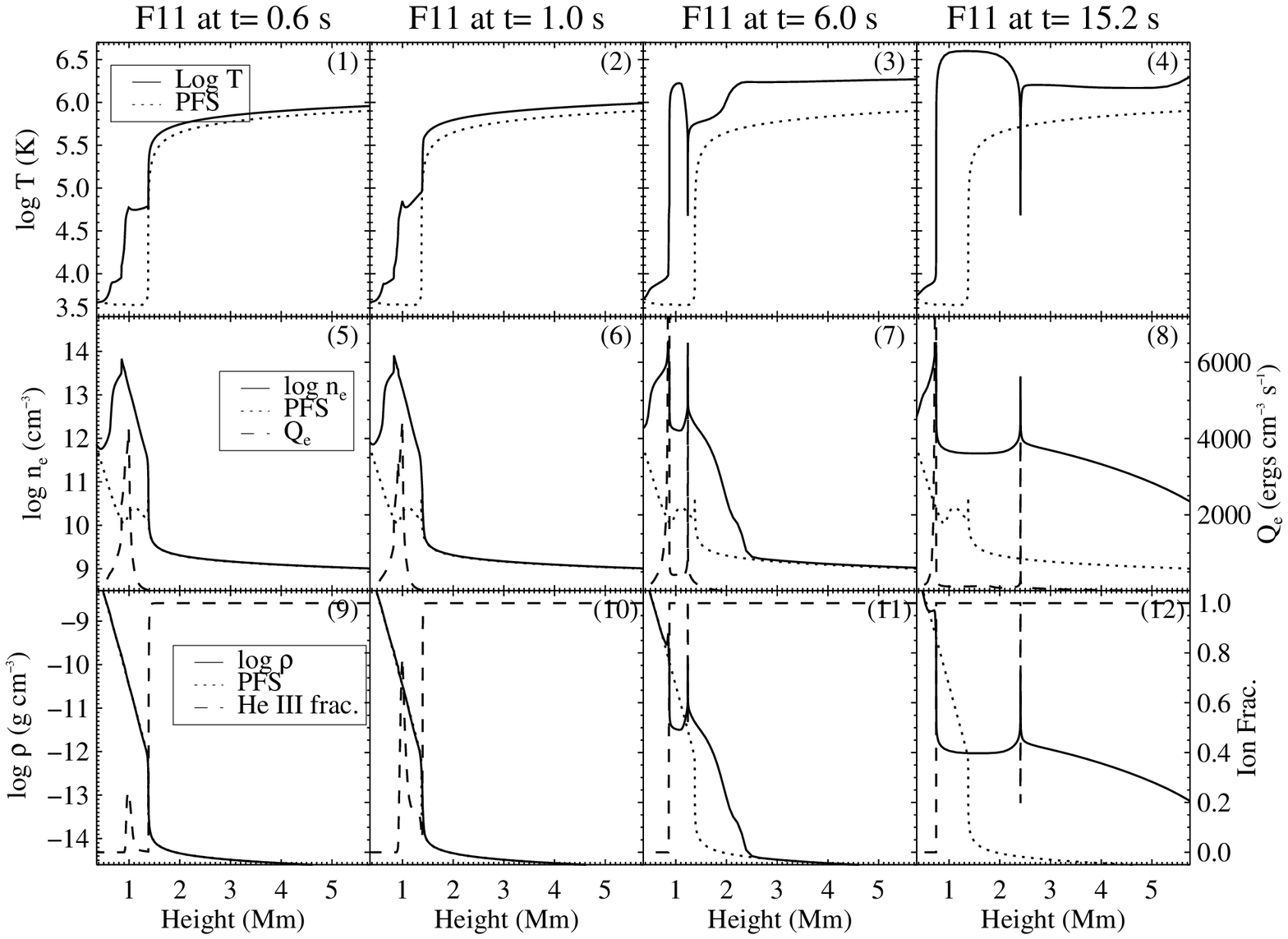}
    \caption{The solar atmosphere at four times during the F11 flare.
    The quantities plotted are identical to Fig.~\ref{fig:f10gen}
    except that the He \textsc{iii} fraction is plotted in the
bottom row rather the hydrogen ionization fraction.}
    \label{fig:f11atmos}
\end{figure}

\begin{figure}
    \plotone{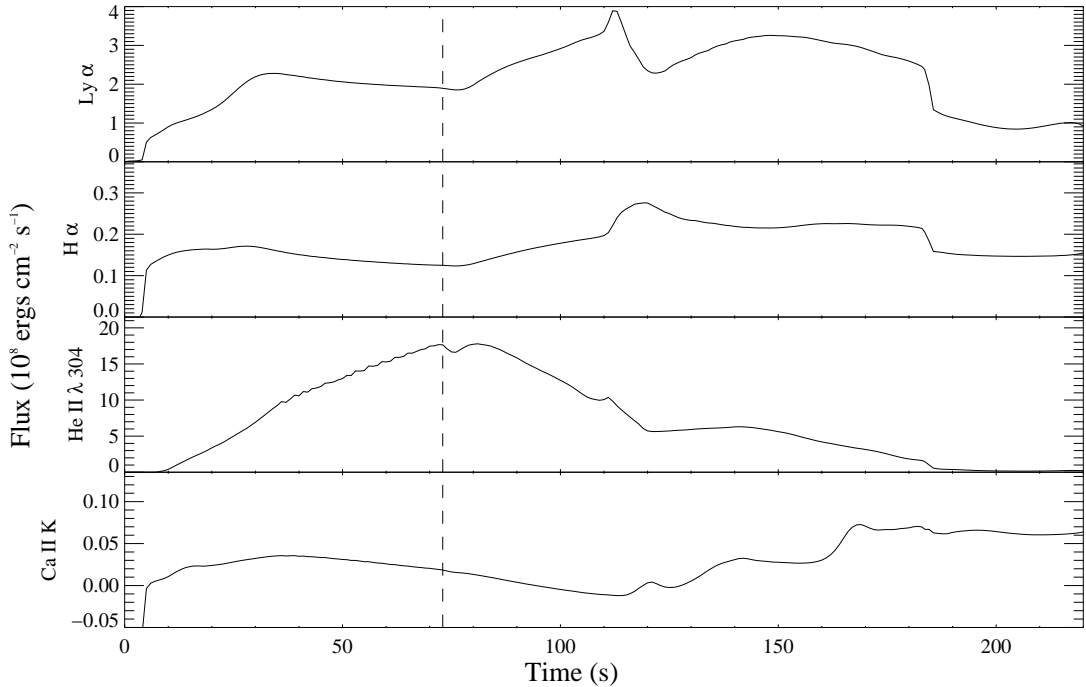}
    \caption{Light curves for Ly$\alpha$, H$\alpha$, He~\textsc{ii}~$\lambda$~304
    and Ca~\textsc{ii} K during the F10 flare.  The preflare
emission has been subtracted.  The line flux for Ca~\textsc{ii} K
includes only the central reversal.  The explosive phase begins
at 73~s and is marked with a dashed line. }
    \label{fig:lc}
\end{figure}

As material evaporates from the chromosphere into higher
temperature regions, many emission lines brighten dramatically.
Light curves showing these increases in flux during the F10 flare
are plotted in Figure~\ref{fig:lc}. Ly$\alpha$ quickly brightens
as lower chromospheric material is heated to $\sim$25,000 K. By
30~s, much of the plasma in the lower transition region has been
further heated and Ly$\alpha$ emission slowly begins to decrease.
The He~\textsc{ii} $\lambda$304 line continually grows from the
beginning of the flare as progressively more material is heated
to its temperature of formation ($\sim$50,000 K) until the onset
of the explosive phase. After the explosive phase begins, the
expansion of the high temperature bubble ionizes most of the
He~\textsc{ii}, and the intensity of He~\textsc{ii} $\lambda$304
decreases (see panels 2 and 3 of Figure~\ref{fig:f10exp}). At the
onset of the explosive phase a downward moving ``condensation''
wave is created and raises the plasma density in the lower
transition region resulting in another increase in Ly$\alpha$
emission. The peak occurring at 112 s is due to emission from the
high density, lower temperature boundary of the expanding bubble
(see panel 3 of Figure~\ref{fig:f10exp}), and the sharp decrease
in intensity occurring at 185 s is the result of the bubble
passing through the boundary of the flux tube. Ca~\textsc{ii} K
does not respond to the explosive phase until the condensation
begins to increase the chromospheric density at about 115 s. At
166~s, a large, downward-moving wave passes into the lower
chromosphere increasing the density and causing Ca~\textsc{ii} K
to brighten (compare the location of the transition region in
panels 3 and 4 of Figure~\ref{fig:f10exp}).

\begin{figure}
    \plotone{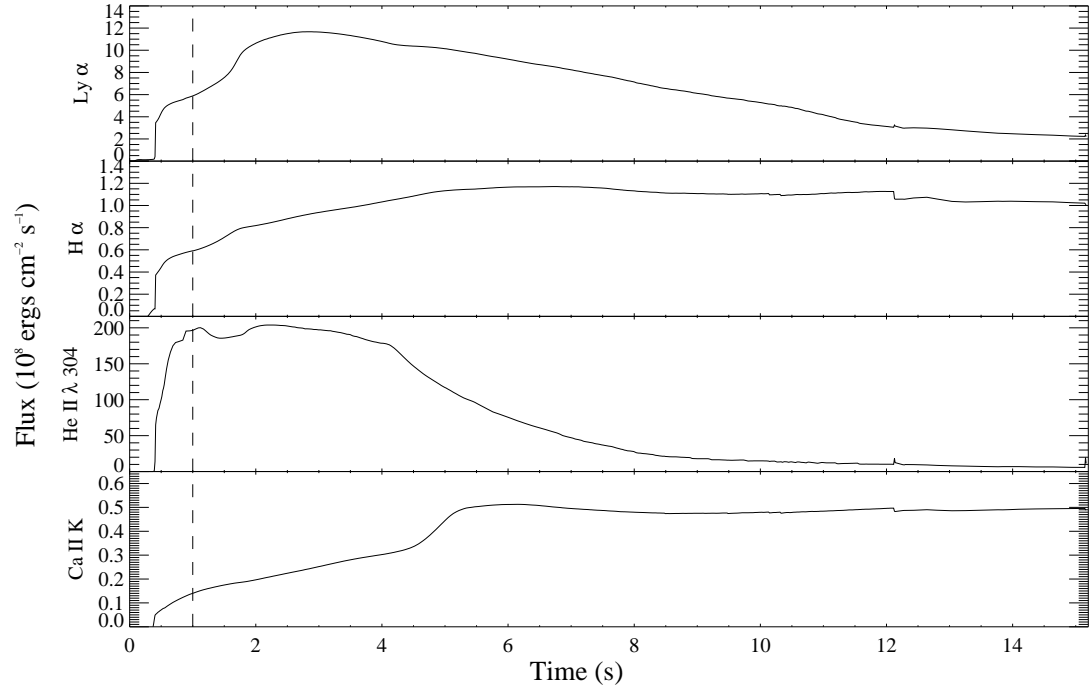}
    \caption{Light curves for Ly$\alpha$, H$\alpha$,
He~\textsc{ii}~$\lambda$~304 and Ca~\textsc{ii} K during the
F11 flare.  The notation is identical to Fig.~\ref{fig:lc}. The
explosive phase begins at 1~s and is marked with a dashed line.}
    \label{fig:f11lc}
\end{figure}

Figure~\ref{fig:f11lc} shows light curves for the same emission
lines during the F11 flare.  As in F10, the lines rapidly
brighten in response to the flare heating.  After 0.5 s, the
region of beam impact has been heated past 40,000 K, and the rate
of intensity increase slows for Ly$\alpha$ and H$\alpha$.  At
2.5~s, the explosive bubble begins to move to areas of lower
density and the Ly$\alpha$ and He~\textsc{ii}~$\lambda$304
intensities slowly decrease. At 5.5~s, a condensation front
passes into the lower chromosphere, increasing the density and
causing Ca~\textsc{ii} K to brighten.

\begin{figure}
    \plotone{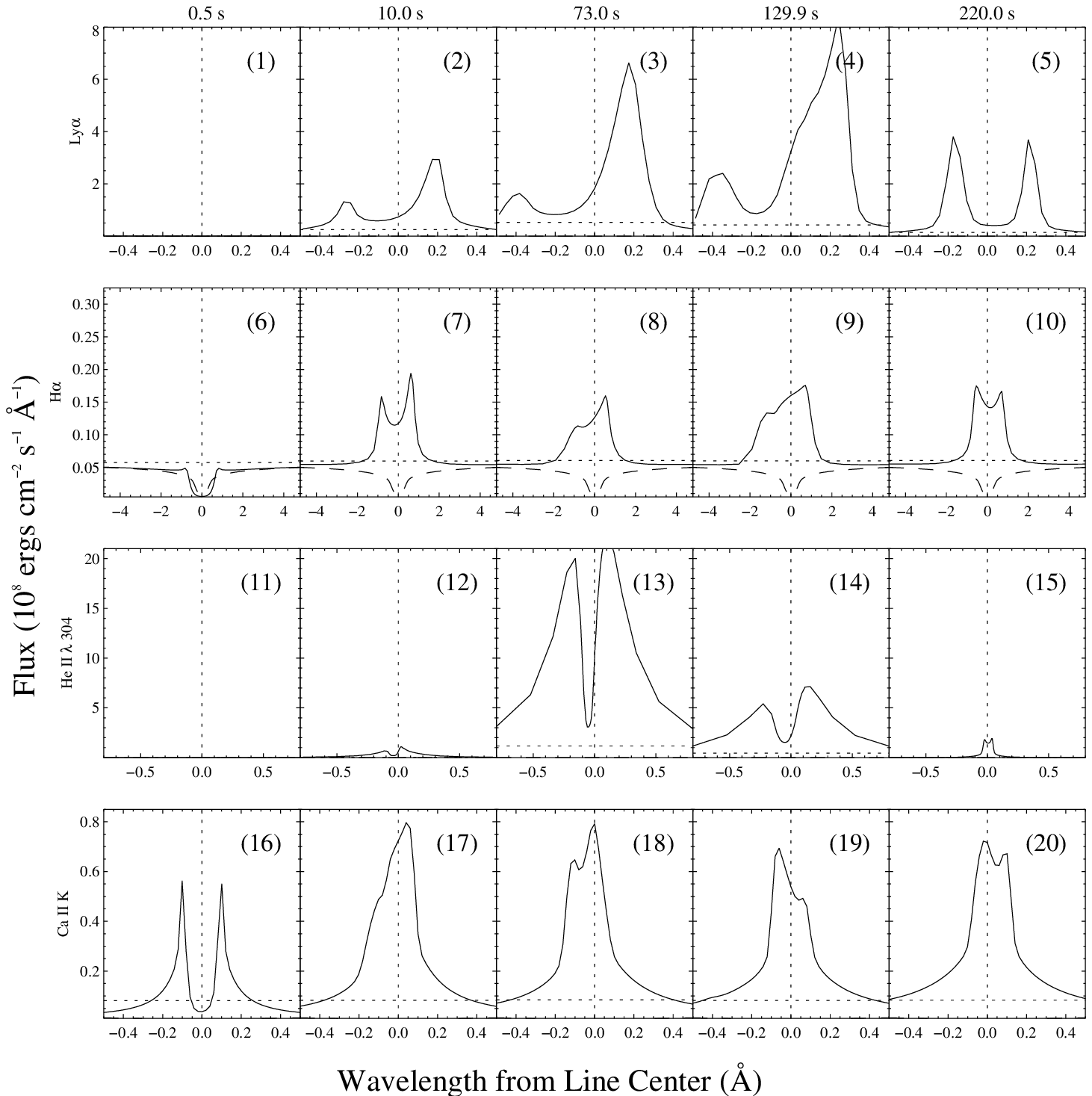}
    \caption{Line profiles for Ly$\alpha$, H$\alpha$,
He~\textsc{ii}~$\lambda$~304 and Ca~\textsc{ii}~K at five times
during the F10 flare (times are indicated at the top of each
column). The dotted lines indicate the level of the continuum and
line center, while the dashed line is the preflare line profile.
The emission in panels 1 and 11 is too small to be seen on this
scale.}
    \label{fig:lprofs}
\end{figure}

Mass motions during the flare cause significant Doppler shifts
and introduce large asymmetries in the line profiles.
Figure~\ref{fig:lprofs} shows profiles for the same emission
lines, at successive times during F10. In panels 2 -- 4, the red
wing of Ly$\alpha$ is strongly enhanced.  Panels 7 -- 9 show
H$\alpha$ to be weakly red shifted (average Doppler shift of 2
km s$^{-1}$).  These indicate chromospheric condensation (mass
motion downward toward the photosphere). However, the line center
of He~\textsc{ii} $\lambda$304 shows the opposite effect (central
reversal in panel 13), with a blue shift exhibiting chromospheric
evaporation. The line center of He~\textsc{ii} $\lambda$304 is
formed higher in the atmosphere --- where the velocity field is
stronger --- than the other lines shown and exhibits a peak
upward velocity of $\sim$120 km s$^{-1}$.  A similar blue shift
has been observed in several solar flares \citep[and references
therein]{s97,b03}. In particular, \citet{b03} observed a velocity
of $\sim$140 km s$^{-1}$ early in the impulsive phase of a
moderate flare using an O~\textsc{iii} line which forms at a
similar temperature to He~\textsc{ii} $\lambda$304.

\begin{figure}
    \plotone{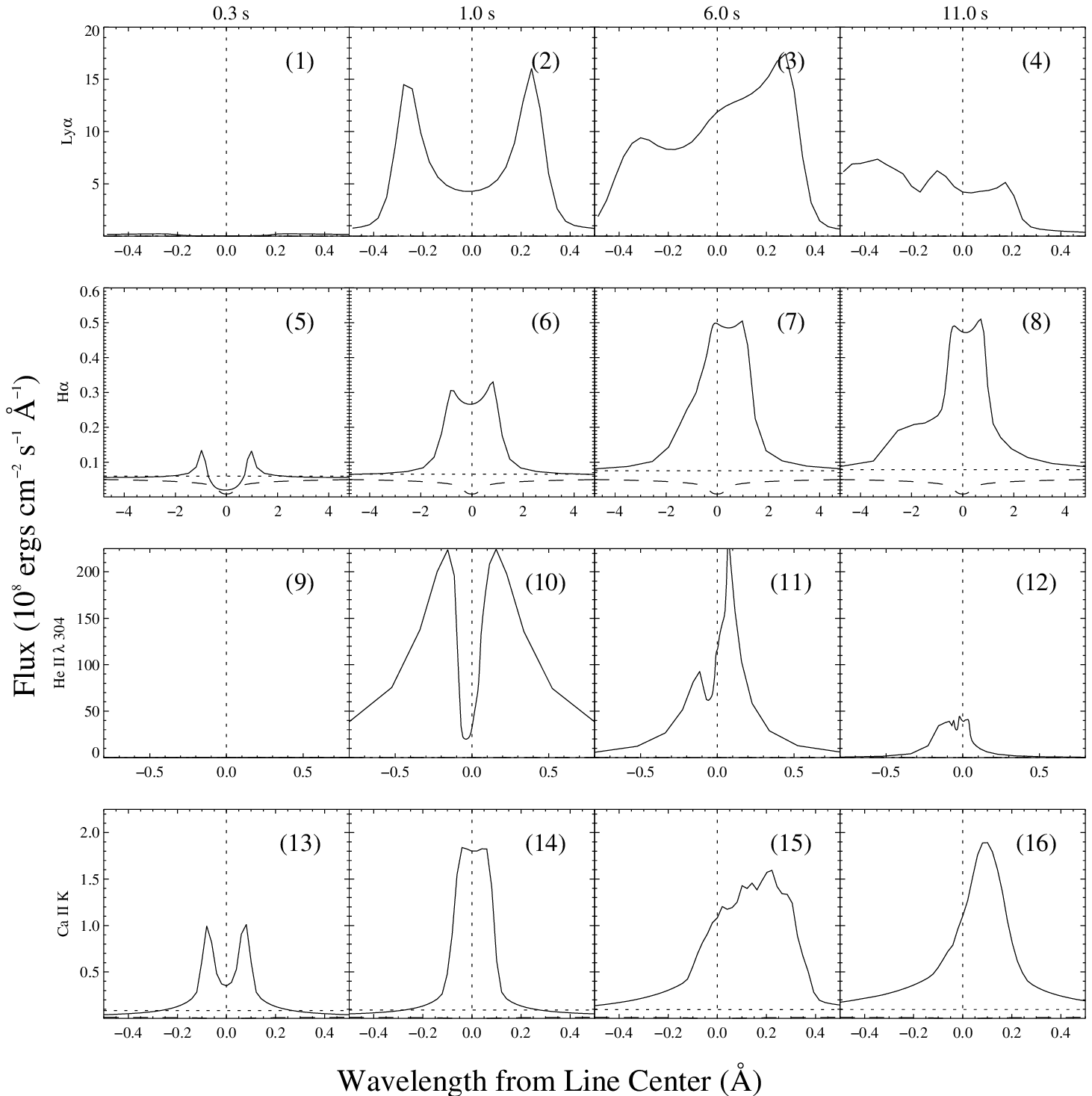}
    \caption{Line profiles for Ly$\alpha$, H$\alpha$,
He~\textsc{ii}~$\lambda$~304 and Ca~\textsc{ii} K at four times
during the F11 flare (times are indicated at the top of each
column). The dotted lines indicate the level of the continuum and
line center, while the dashed line is the preflare line profile.
The emission in panels 1 and 9 is too small to be seen on this
scale.}
    \label{fig:f11profs}
\end{figure}

Figure~\ref{fig:f11profs} shows similar plots for line emission
during the F11 flare.  At 0.3 s both H$\alpha$ and Ca~\textsc{ii}
K have switched from absorption to emission profiles, but neither
Ly$\alpha$ nor He~\textsc{ii} $\lambda$304 has significantly
brightened. However, by 1.0 s the intensity of all these lines
has markedly increased. In panel 10, He~\textsc{ii} $\lambda$304
has a blue shift corresponding to a velocity of $\sim$ 60 km
s$^{-1}$. At 6.0 s there are red shifts present in Ca~\textsc{ii}
K and H$\alpha$ indicating motion towards the photosphere. The
region of H$\alpha$ formation reaches a peak downward velocity of
about 40 km s$^{-1}$ which is consistent with velocities reported
by \citet{ik84}. The profiles of Ly$\alpha$ and He~\textsc{ii}
$\lambda$304 are very asymmetrical since they are formed in
regions with both strong upward and downward directed waves.  It
is, therefore, not possible to determine a single velocity from
these profiles.  In the last column, Ly$\alpha$ and
He~\textsc{ii}~$\lambda$304 show enhanced blue wings, but the
Ca~\textsc{ii}~K line is red-shifted indicating a downward-moving
wave in the lower chromosphere and an upward-directed wave in the
upper chromosphere and transition region.

Clearly, the detailed line profiles in the simulations depend
strongly on the flare dynamics and evolve significantly during the
flare.  To interpret the line profiles, we follow \cite{cs94} and
write the formal solution of the transfer equation for the
emergent intensity as
\begin{equation}
I_\nu(0) =\frac{1}{\mu} \int_{\tau_{\nu}} S_\nu e^{-\tau_\nu/\mu}
d\tau_\nu = \frac{1}{\mu} \int_{z} S_\nu \,
\frac{\chi_\nu}{\tau_\nu} \, \tau_\nu e^{-\tau_\nu/\mu} dz
\end{equation}
where $\chi_\nu$ is the linear extinction coefficient, defined as
the product of the density of emitting particles with their cross
section. The integrand in this equation is known as the
contribution function and indicates how much emission originates
from a height, $z$. We divide the contribution function into
three physically meaningful terms: $\chi_\nu/\tau_\nu$ is large
where there are many emitting particles at small optical depth,
which is a situation that often arises in the presence of strong
velocity gradients. $S_\nu$ is the source function, defined as
the ratio of the emissivity to the extinction coefficient. For a
given line $S_\nu$ is independent of frequency due to the
assumption of complete redistribution. The last term is $\tau_\nu
e^{-\tau_\nu/\mu}$ which represents the attenuation caused by
optical depth and peaks at $\tau_\nu=1$.

\begin{figure}
    \plotone{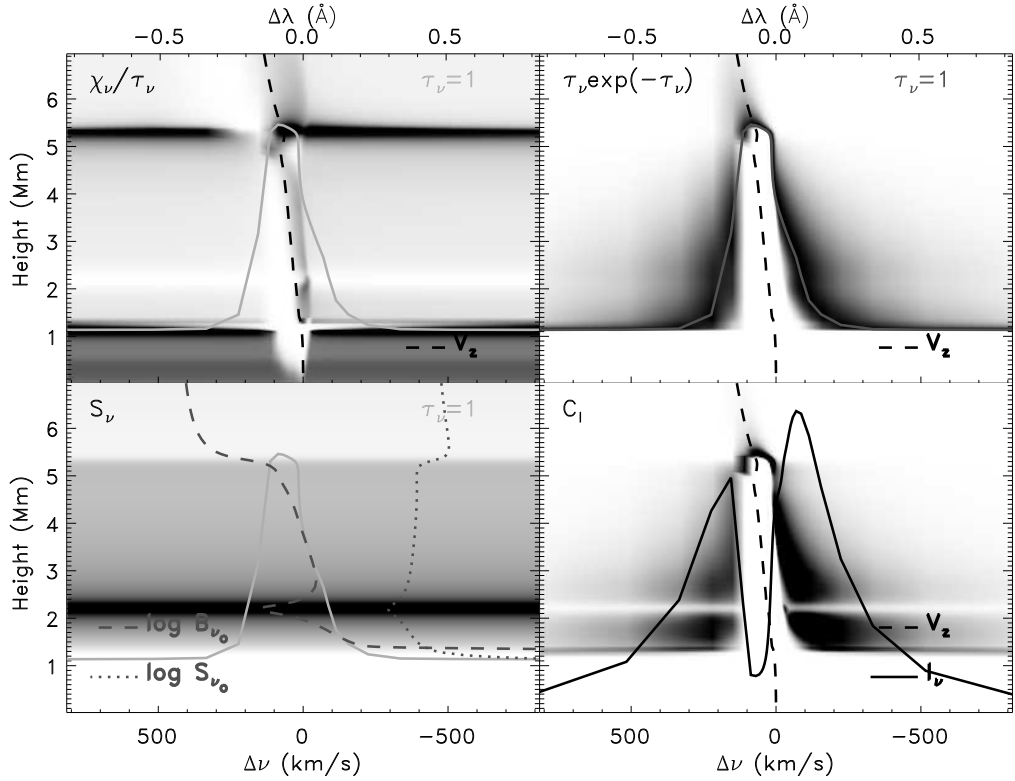}
    \caption{Components of the contribution function for the
He~\textsc{ii} $\lambda$304 line at 73.0 s from the start of the
F10 flare, are plotted in inverse gray scale (high intensity is
darker). The vertical axis is atmospheric height, and the
horizontal axis is frequency from line center in units of km
s$^{-1}$.  The components plotted are $\chi_\nu/\tau_\nu$ (upper
left), $\tau_\nu e^{-\tau_\nu/\mu}$ (upper right), $S_\nu$ (lower
left), and the contribution function, which is the product of the
previous three (lower right).  For reference, the line profile
has been scaled and plotted in the lower-right panel.  The source
function and Planck function at line center are plotted in the
lower left panel to show the extent to which the atmosphere
diverges from LTE.  In addition, lines indicating $\tau_\nu =1$
are plotted in the top two and bottom left panels, and the plasma
velocity is plotted in the top two and lower right panels.
Positive velocity indicates material moving outward, i.e. in the
direction of the corona and correspond to blue shifts in the line
profile (see $\lambda$ scale on top axis).}
    \label{fig:cfheii}
\end{figure}

As an example, we consider the contribution function for the
He~\textsc{ii} $\lambda$304 line at 73 s during the F10 flare.
This is plotted in Figure~\ref{fig:cfheii} in inverse gray scale
(darker shades means higher intensity).  The line profile, shown
in the bottom right panel, has two large peaks in the wings on
either side of line center. The upper right panel ($\tau_\nu
e^{-\tau_\nu/\mu}$) shows that, in the wings, $\tau_\nu =1$ in the
range of $1.1 - 4$ Mm.  In this range the source function and
$\chi_\nu/\tau_\nu$ are both large leading to bright emission in
the wings from this atmospheric range.  At line center, however,
$\tau_\nu =1$ in a relatively small region around 5.2 Mm. At this
height the source function is small resulting in a smaller
contribution function and less emission.

\begin{figure}
    \plotone{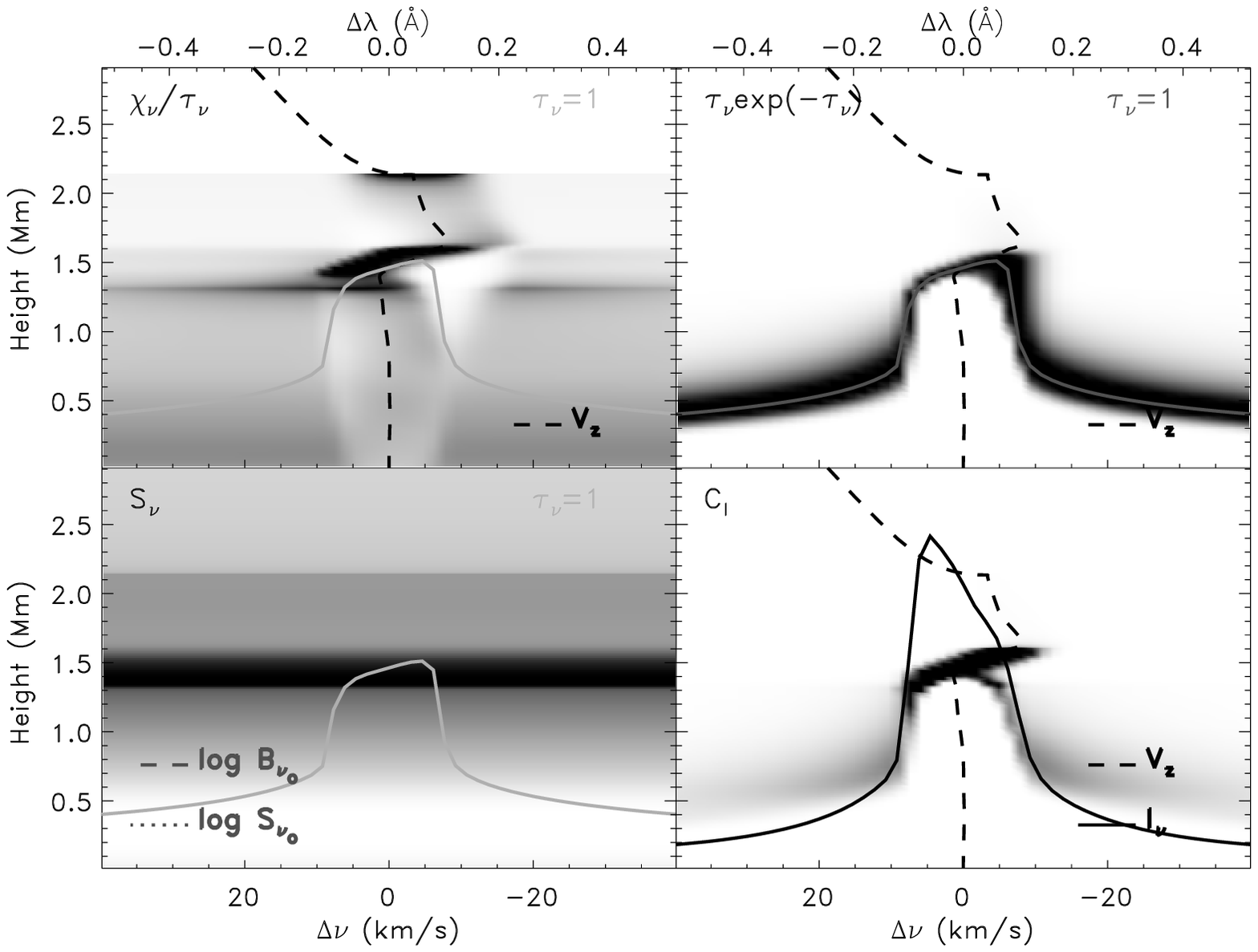}
    \caption{Components to the contribution function for the
Ca~\textsc{ii} K line at 130.0 s from the start of the F10 flare.
The plot is in inverse gray scale.  The notation is identical to
Fig.~\ref{fig:cfheii}.}
    \label{fig:cfcaii}
\end{figure}

The He~\textsc{ii} $\lambda$304 line is formed in the transition
region and is made complicated by the strong velocity gradients
occurring there.  In the lower chromosphere the velocity field is
much smaller.  As an example of emission resulting from this
region, we consider, in Figure~\ref{fig:cfcaii}, the contribution
function for Ca~\textsc{ii}~K at 130 s after the start of the F10
flare. For line center, $\tau_\nu =1$ at a height of about 1.5 Mm.
In this region $\chi_\nu/\tau_\nu$ and $S_\nu$ are also at their
maximums indicating that most of the emission originates from this
region. The wings are formed in the region of $0.3 - 0.5$ Mm where
$\tau_\nu e^{-\tau_\nu/\mu}$ is at a maximum. However, the source
function and $\chi_\nu/\tau_\nu$ are smaller than at line center,
so there is less emission in the wings.

\begin{figure}
    \plotone{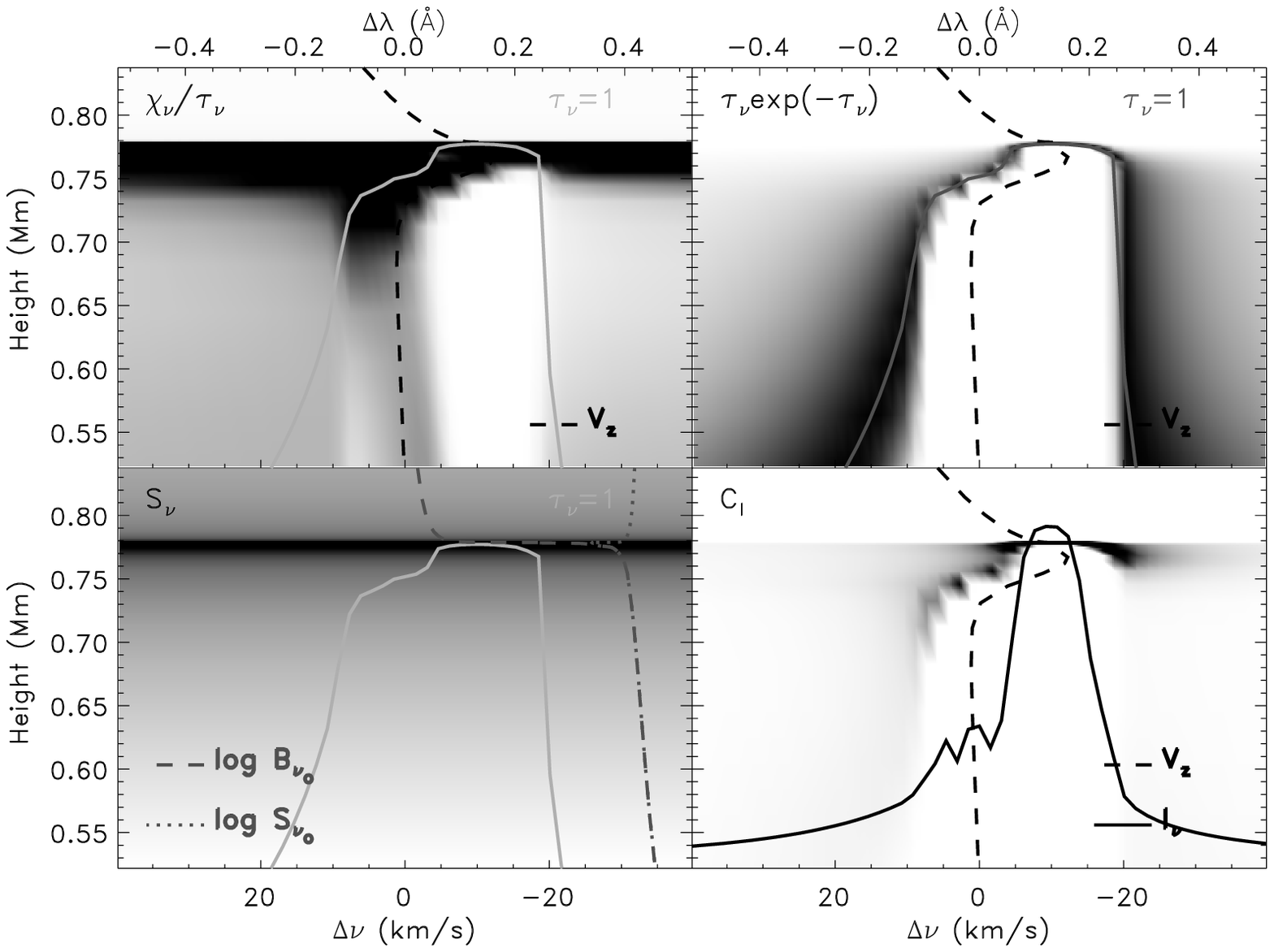}
    \caption{Components to the contribution function for the
Ca~\textsc{ii} K line at 11.0 s from the start of the F11 flare.
The plot is in inverse gray scale.  The notation is identical to
Fig.~\ref{fig:cfheii}.}
    \label{fig:f11cfcaii}
\end{figure}

The line asymmetry of Ca~\textsc{ii}~K is more pronounced in the
F11 flare (compare the last rows of Figure~\ref{fig:lprofs} and
\ref{fig:f11profs}). Figure~\ref{fig:f11cfcaii} shows the
Ca~\textsc{ii}~K contribution function at 11.0 s after the start
of F11. Most of the emission originates from an area with a strong
downward directed condensation wave. Comparison of this figure
with Figure~\ref{fig:cfcaii} reveals that Ca~\textsc{ii}~K is
formed deeper and in a narrower region in F11 than in F10, with a
majority of the emission coming from a height of 0.78 Mm. During
the F11 explosive phase the beam energy is deposited at lower
height than in F10. This has the effect of pushing the transition
region to lower heights and compressing the chromosphere relative
to F10.

\subsection{Continuum Emission}
\begin{figure}
    \plotone{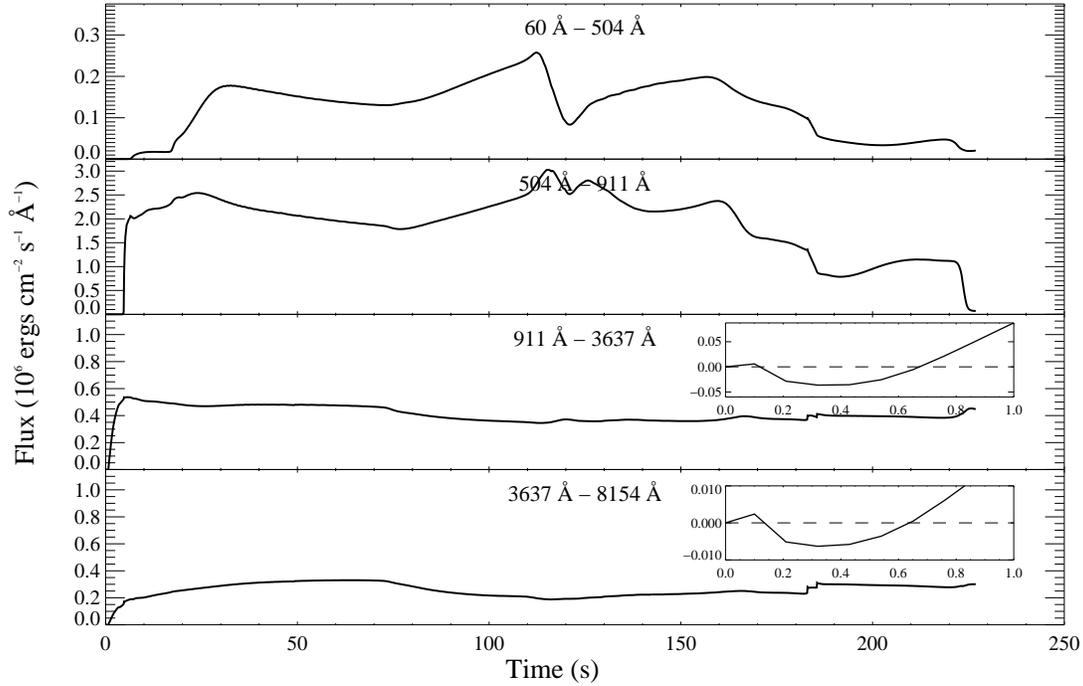}
    \caption{Light curves are shown for four continuum regions
during the F10 flare.  The preflare continuum has been
subtracted. The insets in the bottom two panels show the first
second of the flare, and illustrate the initial continuum
dimming. The major contributors to emission in each bin are
He~\textsc{i} and He~\textsc{ii} continua in the top bin; Lyman
and He~\textsc{ii} continua in the second bin; Balmer,
He~\textsc{i}, and Ca~\textsc{ii} continua in the third bin; and
Paschen and He~\textsc{i} continua in the bottom bin.}
    \label{fig:clc}
\end{figure}

\begin{figure}
    \plotone{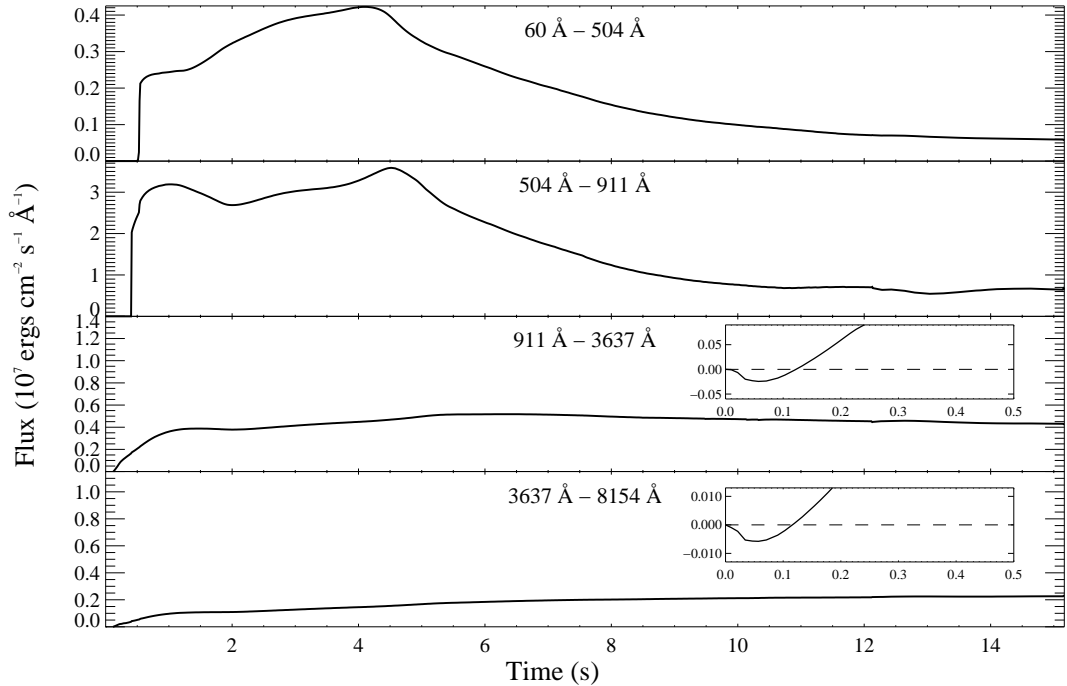}
    \caption{Light curves are shown for four continuum regions
during the F11 flare. The notation is identical to
Fig.~\ref{fig:clc}.  Insets show the first 0.5~s of the flare and
illustrate the initial continuum dimming.}
    \label{fig:f11clc}
\end{figure}

Observations of solar white light flares show enhanced optical
continuum emission that is temporally correlated with the
impulsive phase of the flare \citep{h92}. Our flare simulations
also show dramatically increased continuum emission including the
enhanced Balmer and Paschen jumps which are ubiquitous in type I
white light flares \citep{m03}. Figures~\ref{fig:clc} and
\ref{fig:f11clc} show continuum light curves for four spectral
regions during F10 and F11 respectively. The Lyman continuum is
very enhanced, reaching thousands of times its preflare value.
This compares well with the increased Lyman continuum observed in
an X-class solar flare by \citet{l04}. Since the preflare
intensity is much higher in the optical region, the relative
increase of optical emission is moderate. At 5000 \AA\, the
optical continuum reaches a peak increase of about 5\% for F10
and 30\% for F11. This result is very similar to observations.
\citet{cd05} observed, for a white light flare with a beam
strength of $1.0 \times 10^{10}$ ergs cm$^{-2}$ s$^{-1}$, a
continuum enhancement of $\sim4$ -- 5\%. In our simulations, both
the Balmer and Paschen continua initially decrease in intensity
(see the insets in Fig.~\ref{fig:clc}). This is discussed in
AH99, and is due to the non-thermal ionization resulting from the
electron beam.  The beam raises collisional rates in the upper
chromosphere causing a higher population density of excited
states of hydrogen.  This increases the probability that Balmer
and higher order hydrogen continuum photons, which previously
would have escaped, are absorbed, and results in a decrease in
the continuum intensity.

\begin{figure}
   \plotone{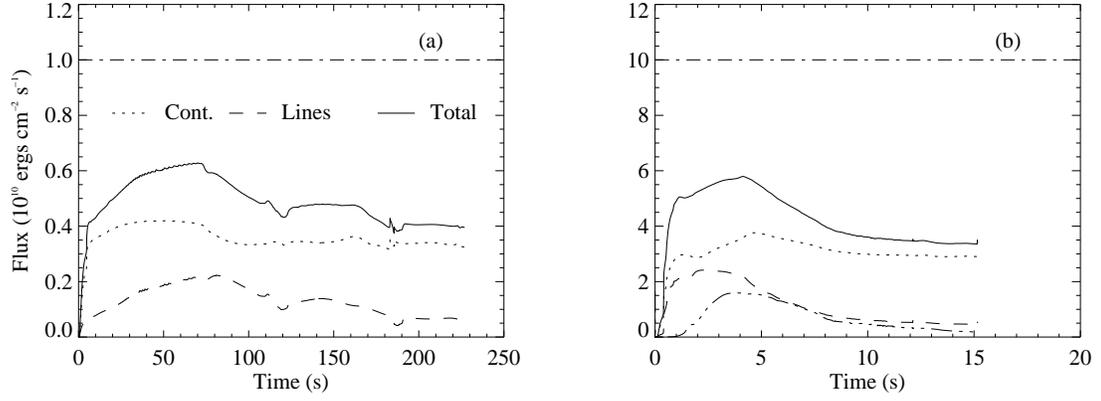}
   \epsscale{1}
   \caption{Light curves of the total line and continuum fluxes
are given for F10 (a) and F11 (b).  The dotted and dashed lines
indicate totals over continua and lines respectively, and the
solid line is the sum of lines and continua. In each case, the
preflare emission has been subtracted so that the plots represent
the emission resulting from the flare heating.  The dot-dashed
line indicates the level of non-thermal beam heating, and the
dashed-dot-dot line in panel (b) is the XEUV heating for F11. For
the F10 flare, the XEUV heating is too small to be seen on this
scale.}
   \label{fig:linecont}
\end{figure}

It is interesting to compare the total fluxes in the lines and
continua, as shown in Figure~\ref{fig:linecont}.  The
beam and XEUV heating is also indicated. Despite differences in
dynamics and duration, in both the F10 and F11 flares the
continuum carries 78\% of the total radiation and the lines carry
the remaining 22\%.

\subsection{Comparison to the AH99 Models}
As has been noted, these models differ from those of AH99 by
including more realistic electron beam heating and greatly
increased XEUV heating. This leads to a number of observational
differences. Our electron beam penetrates deeper (see
Fig.~\ref{fig:qe}) and more directly heats the lower atmosphere.
This results in a larger emitting region for the hydrogen Balmer
lines (compare Fig.~\ref{fig:f10gen} with Fig.~3 of AH99) and
more intense Balmer emission. This is especially important for
the Balmer continuum, which is responsible for radiative
backwarming below the temperature minimum region, where white
light continuum emission is produced \citep{m90}. Since our
models have more intense Balmer continuum, they produce more
backwarming, a hotter flaring photosphere, and more intense white
light emission than the models of AH99.  For the F11 flare, AH99
found a 4\% increase in white light emission, compared to the
30\% increase we obtain in our F11 model.  Another difference
between our models and AH99 is the amount of
He~\textsc{ii}~$\lambda$304 emission.  Our spatially extended
beam deposition produces a pronounced helium-emitting plateau in
the transition region during F11, and results in an order of
magnitude brightening of the He~\textsc{ii}~$\lambda$304 line
compared to AH99.

Despite the brighter emission in our simulations, the velocities
and many line profiles are similar to those found in AH99.  For
example, in the F11 flare AH99 obtain a downward chromospheric
condensation velocity of $\sim40$ km s$^{-1}$, which is
comparable to our result. Also, the \ion{Ca}{2}~K line
profile at 73 s in our F10 flare (panel 18 of
Fig.~\ref{fig:lprofs}) is very similar to the profile calculated
at 30~s in the F10 model of AH99 (panel 6 of Fig.~14 of AH99). As
noted earlier our simulations evolve more slowly than those of
AH99, so these line profiles occur near the beginning of the
explosive phase in each simulation.

\section{Limitations and Plans for Future Models}
We have created detailed models of the radiation and dynamics
occurring during a moderate and a strong solar flare and have
taken considerable effort to make them as accurate as possible.
However, there are a number of limitations to these models. The
atmosphere is assumed to be one-dimensional, plane-parallel and
aligned with the magnetic field. In assuming a one-dimensional
atmosphere, the possibility that radiation can escape the flux
tube through the sides is neglected. The assumption of a circular
loop geometry neglects the more tapered structure of true
magnetic field loops and will affect the calculation of the
thermal XEUV heating. Ions and electrons are assumed to have the
same temperature.  We also have assumed complete redistribution
in calculating all transitions, and we have implicitly assumed
that the gas pressure is lower than the magnetic pressure so that
the gas is confined to the flux tube and MHD effects can be
neglected.

We are in the process of making a number of improvements to the
models. The models shown in this paper represent a ``general''
flare, i.e. the electron beam parameters are assumed to remain
constant throughout the flare.  RHESSI hard X-ray spectral
observations have sufficient time resolution to see temporal
variations in the electron beam on the order of a few seconds
allowing the evolving electron beam parameters to be inferred.
Thus, simulations of specific flares may be constructed, and the
calculated emission profiles can be directly compared to
observations of lines and continua for that specific flare.  A
forthcoming paper will present results for a simulation of the
large X-class 23 July 2002 solar flare.

\section{Conclusions}
We have constructed radiative hydrodynamic simulations of the
flaring solar atmosphere and have used them to explore the
flare-induced optical and UV emission.  These models extend the
work presented in AH99 by using more realistic electron beam
deposition rates and by incorporating backwarming from a large
number of XEUV lines.  As in AH99, we find that the
impulsive flare naturally divides into two phases, an initial
gentle phase followed by a period of explosive increases in
temperature and pressure.  The explosive wave front creates a high
temperature ``bubble.''  The bubble expands until the wave passes
through the boundary of the flux tube, at which point the
atmosphere attains a steady state with a hotter corona and deeper
transition region than in the preflare atmosphere.

Both moderate and strong flares show large increases in line and
continuum emission. As chromospheric material is heated to
transition region temperatures, the He~\textsc{ii}~$\lambda$304
line is especially enhanced reaching values of many thousands of
times its preflare level.  The optical continuum also brightens
showing a peak enhancement of 5\% for F10 (comparable to
observed enhancements) and 30\% for F11. The continuum dominates
the emitted radiation, carrying 78\% of the total emission for
both F10 and
F11. Chromospheric evaporation causes blue-shifts in transition
region lines corresponding to a peak velocity of
$\sim120$ km s$^{-1}$. Chromospheric condensation increases
the density in the chromosphere causing Ca~\textsc{ii}~K and
H$\alpha$ to significantly brighten and resulting in an H$\alpha$
red shift of $\sim40$ km s$^{-1}$.

We incorporated the large number of transitions in the ATOMDB
database to more accurately model XEUV backwarming of the lower
atmosphere.  We find that the XEUV heating is an order of
magnitude larger than the soft X-ray heating of AH99.
Nevertheless, the level of heating is still too small to
significantly contribute to the flare dynamics. By comparing
these models with those in AH99 we find that heating from a
more realistic electron energy distribution leads to slower flare
evolution, and more intense brightening of emission lines and
continua.  The larger Balmer continuum present in our simulations
is especially important because it radiatively backwarms the
lower solar atmosphere, enhanced white light emission.

\acknowledgments
This work has been partially funded by NSF grant AST02-05875 and
HST grants AR-10312 and GO-8613. The computations presented here
were carried out on the Astronomy Condor Network at the
University of Washington, and we would especially like to thank
John Bochanski for contributing many hours of computer time.
%--------------------------BIBLIOGRAPHY--------------------------

%--------------------------Tables---------------------------------
\clearpage
\begin{deluxetable}{lcccccccc}
%\tabletypesize{\scriptsize}
\tablewidth{0pt}
\tablecaption{Bound-Bound Transitions \label{table:bb}}
\tablecolumns{6}
\tablehead{
\colhead{Atom} & \colhead{$\lambda_{ij}$ (\AA)} &
\colhead{Transition} &\colhead{Atom} & \colhead{$\lambda_{ij}$
(\AA)} & \colhead{Transition} \\ }\startdata
H  {\sc i} & 1215.70 & Ly$\alpha$ &
   Ca {\sc ii} & 3933.65 & $2s$ -- $2p_{3}$ \\
& 1025.75 & Ly$\beta$ &
    & 8662.16 & $2d_{3}$ -- $2p_{1}$ \\
& 972.56 & Ly$\gamma$ &
    & 8498.01 & $2d_{3}$ -- $2p_{3}$ \\
& 949.77 & Ly$\delta$ &
    & 8542.05 & $2d_{5}$ -- $2p_{3}$ \\
& 6562.96 & H$\alpha$ &
   He \sc{i} & 625.58 & $1s^{2}$ $^1S_{0}$ -- $1s$ $2s$ $^3S_{1}$\\
& 4861.50 & H$\beta$ &
    & 601.42 & $1s^{2}$ $^1S_{0}$ -- $1s$ $2s$ $^1S_{0}$ \\
& 4340.62 & H$\gamma$ &
    & 10830.29 & $1s$ $2s$ $^3S_{1}$ -- $1s$ $2p$ $^{3}P^{0}_{4}$\\
& 18752.27 & P$\alpha$ &
    & 584.35 & $1s^{2}$ $^1S_{0}$ -- $1s$ $2p$ $^{1}P^{0}_{1}$ \\
& 12818.86 & P$\beta$ &
    & 20580.82 & $1s$ $2s$ $^1S_{0}$ -- $1s$ $2p$ $^{1}P^{0}_{1}$\\
& 40513.47 & B$\alpha$ &
    He \sc{ii} & 303.79 & $1s$ $^2S_{1/2}$ -- $1s$ $2p$ $^{2}S_{1/2}$ \\
Ca \sc{ii} & 3968.46 & $2s$ -- $2p_{1}$ &
    & 303.78 & $1s$ $^2S_{1/2}$ -- $2p$ $^{2}P^{0}_{2}$
\enddata
\end{deluxetable}

\begin{deluxetable}{lcccccccc}
\tablewidth{0pt} \tablecaption{Bound-Free Transitions
\label{table:bf}} \tablecolumns{6} \tablehead{ \colhead{Atom} &
\colhead{$\lambda_{ic}$ (\AA)} & \colhead{Initial State} &
\colhead{Atom} & \colhead{$\lambda_{ic}$ (\AA)} & \colhead{Initial State} \\ }
\startdata
H {\sc i} & 911.12 & n=1 &
   Ca {\sc ii} & 1421.04 & $4p ^2P^0_{3/2}$ \\
H {\sc i} & 3635.67 & n=2 &
   He {\sc i} & 503.98 & $1s^2$ $^{1}S_0$ \\
H {\sc i} & 8151.31 & n=3 &
   He {\sc i} & 2592.02 & $1s$ $2s$ $^{3}S_1$ \\
H {\sc i} & 14419.07 & n=4 &
   He {\sc i} & 3109.80 & $1s$ $2s$  $^{1}S_0$\\
H {\sc i} & 22386.68 & n=5 &
   He {\sc i} & 3407.63 & $1s$ $2p$ $^{3}P^0_4$ \\
Ca {\sc ii} & 1044.00 & $4s ^2S_{1/2}$ &
   He {\sc i} & 3663.37 & $1s$ $2p$ $^{1}P^0_1$ \\
Ca {\sc ii} & 1217.50 &  $3d ^2D_{3/2}$&
   He {\sc ii} & 227.84 & $1s$ $^{2}S_{1/2}$ \\
Ca {\sc ii} & 1218.40 & $3d ^2D_{5/2}$ &
   He {\sc ii} & 911.34 & $2s$ $^{2}S_{1/2}$ \\
Ca {\sc ii} & 1416.55 & $4p ^2P^0_{1/2}$ &
   He {\sc ii} & 911.36 & $2p$ $^{2}P^{0}_{1/2}$
\enddata
\end{deluxetable}

\begin{deluxetable}{lc}
\tablewidth{0pt}
\tablecaption{Soft X-ray EUV Wavelength Bins
\label{table:xbins}}
\tablecolumns{2}
\tablehead{ \colhead{Wavelength Range (\AA$\,$)} & \colhead{Central Wavelength (\AA$\,$)}}
\startdata
       1 -      15 &      11.2 \\
      15 -      25 &      18.6 \\
      25 -      50 &      37.2 \\
      50 -     100 &      72.3 \\
     100 -     150 &      125.5 \\
     150 -     200 &      177.8 \\
     200 -     250 &      224.7 \\
     250 -     300 &      269.2 \\
     300 -     500 &      352.4 \\
     500 -     750 &      608.5 \\
     750 -    1000 &      867.7 \\
    1000 -    1500 &      1163.6 \\
    1500 -    2000 &      1695.4 \\
    2000 -    2500 &      2362.1
\enddata
\end{deluxetable}

%--------------------------Figures--------------------------------
\clearpage
\clearpage
\clearpage

\begin{thebibliography}{}
\bibitem[Abbett(1998)]{a98} Abbett, W. P.\ 1998, Ph.D. thesis, Michigan State Univ.
\bibitem[Abbett \& Hawley(1999)]{ah99} Abbett W. P. \& Hawley S. L. 1999, \apj,521, 906
\bibitem[Brosius(2003)]{b03} Brosius, J.~W.\ 2003, \apj, 586, 1417
\bibitem[Brown(1973)]{b73} Brown, J.C. 1973, Sol. Phys., 31, 143
\bibitem[Carlsson(1986)]{c86} Carlsson, M.\ 1986 Uppsala Astronomical Report No. 33
\bibitem[Carlsson \& Stein(1994)]{cs94} Carlsson, M., \& Stein, R.~F.\ 1994, in Proc. Mini-Workshop on Chromospheric Dynamics, ed. M.~Carlsson (Oslo: Institute of Theoretical Astrophysics), 47
\bibitem[Carlsson \& Stein(1995)]{cs95} Carlsson, M., \& Stein, R.~F.\ 1995, \apjl, 440, L29
\bibitem[Carlsson \& Stein(1997)]{cs97} Carlsson, M.~\& Stein, R.~F.\ 1997, \apj, 481, 500
\bibitem[Chen \& Ding(2005)]{cd05} Chen, Q.~R., \& Ding, M.~D.\ 2005, \apj, 618, 537
\bibitem[Cheng et al.(1983)]{c83} Cheng, C.-C., Oran, E.~S., Doschek, G.~A., Boris, J.~P., \& Mariska, J.~T.\ 1983, \apj, 265, 1090
\bibitem[Cheng et al.(1984)]{c84} Cheng, C.-C., Doschek, G.~A., \& Karpen, J.~T.\ 1984, \apj, 286, 787
\bibitem[Dere et al.(1997)]{d97} Dere, K.~P., Landi, E.,Mason, H.~E., Monsignori
Fossi, B.~C., \& Young, P.~R.\ 1997, \aaps, 125,149
\bibitem[Dorfi \& Drury(1987)]{dd87} Dorfi, E. A., \& Drury, L. O.\ 1987 J. Comput. Phys., 69, 175
\bibitem[Emslie(1978)]{e78} Emslie, A.~G.\ 1978, \apj, 224,241
\bibitem[Emslie(1981)]{e81} Emslie, A.~G.\ 1981, \apj, 245,711
\bibitem[Fisher, Canfield, \& McClymont(1985)]{fcm85} Fisher,G.~H., Canfield, R.~C., \& McClymont, A.~N.\ 1985, \apj,
289, 414
\bibitem[Gan \& Fang (1990)]{gan90} Gan, W. Q., \& Fang, C. 1990, ApJ, 358, 328.
\bibitem[Gustafsson(1973)]{g73} Gustafsson, B.\ 1973 A FORTRAN Program for Calculating ``Continuous'' Absorption
Coefficients of Stellar Atmospheres(Uppsala: Landstingets Vergstader)
\bibitem[Hawley \& Fisher(1994)]{hf94} Hawley, S.~L.~\& Fisher, G.~H.\ 1994, \apj, 426, 387
\bibitem[Henoux \& Nakagawa (1977)]{hen77} Henoux, J. C., \& Nakagawa, Y., 1977, A\&A, 57, 105.
\bibitem[Holman et al.(2003)]{h03}Holman, G.~D., Sui, L., Schwartz, R.~A., \& Emslie, A.~G.\ 2003, \apjl,595, L97
\bibitem[Hudson et al.(1992)]{h92} Hudson, H.~S., Acton, L.~W., Hirayama, T., \& Uchida, Y.\ 1992, \pasj, 44, L77
\bibitem[Ichimoto \& Kurokawa(1984)]{ik84} Ichimoto, K., \& Kurokawa, H.\ 1984, \solphys, 93, 105
\bibitem[Lemaire et al.(2004)]{l04} Lemaire, P., Gouttebroze, P., Vial, J.-C., Curdt, W., Sch{\" u}hle, U., \& Wilhelm, K.\ 2004, \aap, 418, 737
\bibitem[MacNeice et al.(1984)]{m84} MacNeice, P., Burgess, A., McWhirter, R.~W.~P., \& Spicer, D.~S.\ 1984,
  \solphys, 90, 357
\bibitem[Metcalf et al.(1990)]{m90} Metcalf, T.~R., Canfield, R.~C., \& Saba, J.~L.~R.\ 1990, \apj, 365, 391
\bibitem[Metcalf et al.(2003)]{m03} Metcalf, T.~R., Alexander, D., Hudson, H.~S., \& Longcope, D.~W.\ 2003, \apj, 595, 483
\bibitem[Mariska, Emslie, \& Li(1989)]{mel89} Mariska, J.~T., Emslie, A.~G., \& Li, P.\ 1989, \apj, 341, 1067, 359, 524
\bibitem[Nagai \& Emslie(1984)]{ne84} Nagai, F., \& Emslie, A.~G.\ 1984, \apj, 279, 896
\bibitem[Peres et al.(1987)]{p87} Peres, G., Reale, F., Serio, S., \& Pallavicini, R.\ 1987, \apj, 312, 895
\bibitem[Peres \& Reale(1993)]{pr93} Peres, G., \& Reale, F.\ 1993, \aap, 267, 566
\bibitem[Reale \& Peres(1995)]{rp95} Reale, F., \& Peres, G.\ 1995, \aap, 299, 225
\bibitem[Ricchiazzi \& Canfield(1983)]{rc83} Ricchiazzi, P.~J.~\& Canfield, R.~C.\ 1983, \apj, 272, 739
\bibitem[Scharmer \& Carlsson(1985)]{sc85} Scharmer, G. B.,\&  Carlsson, M.\ 1985 J. Comput. Phys., 59, 56
\bibitem[Silva et al.(1997)]{s97} Silva, A.~V.~R., Wang, H., Gary, D.~E., Nitta, N., \& Zirin, H.\ 1997, \apj, 481, 978
\bibitem[Smith et al.(2001)]{sblr01} Smith, R.~K., Brickhouse, N.~S., Liedahl, D.~A., \& Raymond, J.~C.\ 2001, \apjl, 556, L91
\bibitem[Young et al.(2003)]{y03} Young, P.~R., Del Zanna, G., Landi, E., Dere, K.~P., Mason, H.~E., \& Landini, M.\ 2003, \apjs, 144, 135
\end{thebibliography}
\end{document}